\documentclass{ws-ijmpcs}

\begin{document}

\markboth{T. Gilliss \& The MAJORANA Collaboration}
{Recent Results from the MAJORANA DEMONSTRATOR}

\def\MJ{M{\footnotesize{AJORANA}}}
\def\DEM{D{\footnotesize{EMONSTRATOR}}}
\def\onbb{$0\nu\beta\beta$}
\def\tnbb{$2\nu\beta\beta$}
\def\onbbx{$0\nu\beta\beta\chi$}
\def\ssGe{$^{76}$Ge}

%
\catchline{}{}{}{}{}
%

\title{Recent Results from the \MJ~\DEM}

\newcommand{\cs}{$^{,}$}
\newcommand{\unc}{$^{1}$}
\newcommand{\tunl}{$^{2}$}
\newcommand{\uw}{$^{3}$}
\newcommand{\pnnl}{$^{4}$}
\newcommand{\usc}{$^{5}$}
\newcommand{\ornl}{$^{6}$}
\newcommand{\itep}{$^{7}$}
\newcommand{\usd}{$^{8}$}
\newcommand{\mpi}{$^{9}$}
\newcommand{\jinr}{$^{10}$}
\newcommand{\duke}{$^{11}$}
\newcommand{\lbnl}{$^{12}$}
\newcommand{\sdsmt}{$^{13}$}
\newcommand{\lanl}{$^{14}$}
\newcommand{\ut}{$^{15}$}
\newcommand{\ou}{$^{16}$}
\newcommand{\princeton}{$^{17}$}
\newcommand{\ncsu}{$^{18}$}
\newcommand{\MIT}{$^{19}$} 
\newcommand{\blhill}{$^{20}$}
\newcommand{\ttu}{$^{21}$}
\newcommand{\queens}{$^{22}$}
\newcommand{\tum}{$^{23}$}
\newcommand{\clara}{$^{a}$}
\newcommand{\chrisO}{$^{b}$}
\newcommand{\vetter}{$^{c}$}



\author{T~Gilliss\unc\cs\tunl, S~I~Alvis\uw, I~J~Arnquist\pnnl, F~T~Avignone~III\usc\cs\ornl, A~S~Barabash\itep, C~J~Barton\usd, F~E~Bertrand\ornl, T~Bode\mpi, V~Brudanin\jinr, M~Busch\tunl\cs\duke, M~Buuck\uw, T~S~Caldwell\unc\cs\tunl, Y-D~Chan\lbnl, C~D~Christofferson\sdsmt, P~-H~Chu\lanl, C~Cuesta\clara\cs\uw, J~A~Detwiler\uw, C~Dunagan\sdsmt, Yu~Efremenko\ornl\cs\ut, H~Ejiri\ou, S~R~Elliott\lanl, G~K~Giovanetti\princeton, M~P~Green\tunl\cs\ornl\cs\ncsu, J~Gruszko\MIT, I~S~Guinn\uw, V~E~Guiseppe\usc, C~R~Haufe\unc\cs\tunl, L~Hehn\lbnl, R~Henning\unc\cs\tunl, E~W~Hoppe\pnnl, M~A~Howe\unc\cs\tunl, K~J~Keeter\blhill, M~F~Kidd\ttu, S~I~Konovalov\itep, R~T~Kouzes\pnnl, A~M~Lopez\ut, R~D~Martin\queens, R~Massarczyk\lanl, S~J~Meijer\unc\cs\tunl, S~Mertens\mpi\cs\tum, J~Myslik\lbnl, C~O'Shaughnessy\chrisO\cs\unc\cs\tunl, G~Othman\unc\cs\tunl, W~Pettus\uw, A~W~P~Poon\lbnl, D~C~Radford\ornl, J~Rager\unc\cs\tunl, A~L~Reine\unc\cs\tunl, K~Rielage\lanl, R~G~H~Robertson\uw, N~W~Ruof\uw, B~Shanks\ornl, M~Shirchenko\jinr, A~M~Suriano\sdsmt, D~Tedeschi\usc, R~L~Varner\ornl, S~Vasilyev\jinr, K~Vetter\vetter\cs\lbnl, K~Vorren\unc\cs\tunl, B~R~White\lanl, J~F~Wilkerson\unc\cs\tunl\cs\ornl, C~Wiseman\usc, W~Xu\usd, E~Yakushev\jinr, C~-H~Yu\ornl, V~Yumatov\itep, I~Zhitnikov\jinr, and B~X~Zhu\lanl}                                                          
\address{\unc Department of Physics and Astronomy, University of North Carolina, Chapel Hill, NC, USA}
\address{\tunl Triangle Universities Nuclear Laboratory, Durham, NC, USA}
\address{\uw Center for Experimental Nuclear Physics and Astrophysics, and Department of Physics, University of Washington, Seattle, WA, USA}
\address{\pnnl Pacific Northwest National Laboratory, Richland, WA, USA}
\address{\usc Department of Physics and Astronomy, University of South Carolina, Columbia, SC, USA}
\address{\ornl Oak Ridge National Laboratory, Oak Ridge, TN, USA}
\address{\itep National Research Center ``Kurchatov Institute'' Institute for Theoretical and Experimental Physics, Moscow, Russia}
\address{\usd Department of Physics, University of South Dakota, Vermillion, SD, USA} 
\address{\mpi Max-Planck-Institut f\"{u}r Physik, M\"{u}nchen, Germany}
\address{\jinr Joint Institute for Nuclear Research, Dubna, Russia}
\address{\duke Department of Physics, Duke University, Durham, NC, USA}
\address{\lbnl Nuclear Science Division, Lawrence Berkeley National Laboratory, Berkeley, CA, USA}
\address{\sdsmt South Dakota School of Mines and Technology, Rapid City, SD, USA}
\address{\lanl Los Alamos National Laboratory, Los Alamos, NM, USA}
\address{\ut Department of Physics and Astronomy, University of Tennessee, Knoxville, TN, USA}
\address{\ou Research Center for Nuclear Physics, Osaka University, Ibaraki, Osaka, Japan}
\address{\princeton Department of Physics, Princeton University, Princeton, NJ, USA}
\address{\ncsu Department of Physics, North Carolina State University, Raleigh, NC, USA}
\address{\MIT Department of Physics, Massachusetts Institute of Technology, Cambridge, MA, USA}
\address{\blhill Department of Physics, Black Hills State University, Spearfish, SD, USA}
\address{\ttu Tennessee Tech University, Cookeville, TN, USA}
\address{\queens Department of Physics, Engineering Physics and Astronomy, Queen's University, Kingston, ON, Canada} 
\address{\tum Physik Department, Technische Universit\"{a}t, M\"{u}nchen, Germany}
\address{\clara Present Address: Centro de Investigaciones Energ\'{e}ticas, Medioambientales y Tecnol\'{o}gicas, CIEMAT, 28040, Madrid, Spain}
\address{\chrisO Present Address: \lanl}
\address{\vetter Alternate address: Department of Nuclear Engineering, University of California, Berkeley, CA, USA}

\maketitle

\begin{history}
\published{15 January 2018}
\end{history}

\begin{abstract}
The \MJ~Collaboration has completed construction and is now operating an array of high purity Ge detectors searching for neutrinoless double-beta decay (\onbb) in \ssGe. The array, known as the \MJ~\DEM, is comprised of 44 kg of Ge detectors (30 kg enriched to 88\% in \ssGe) installed in an ultra-low background compact shield at the Sanford Underground Research Facility in Lead, South Dakota. The primary goal of the \DEM~is to establish a low-background design that can be scaled to a next-generation tonne-scale experiment. This work reports initial background levels in the \onbb~region of interest. Also presented are recent physics results leveraging P-type point-contact detectors with sub-keV energy thresholds to search for physics beyond the Standard Model; first results from searches for bosonic dark matter, solar axions, Pauli exclusion principle violation, and electron decay have been published. Finally, this work discusses the proposed tonne-scale \ssGe~\onbb~LEGEND experiment.
\end{abstract}

\section{Introduction}	

Experimental searches for neutrinoless double-beta decay (\onbb) are motivated by several questions regarding the nature of neutrinos and their role in the evolution of the universe. The observation of \onbb\ would indicate that lepton number is not conserved-- a key ingredient of thermal leptogenesis-- and that the light neutrinos are Majorana particles\cite{SchechterValle1982}.  

Assuming that \onbb~proceeds through the exchange of a light Majorana neutrino, the transition rate of the decay depends on the effective Majorana neutrino mass $m_{\beta\beta}$ as
\begin{equation}
(T^{0\nu}_{1/2})^{-1} = G^{0\nu} \left| M^{0\nu}\right|^2 \left( \frac{\langle m_{\beta\beta} \rangle}{m_e} \right)^2 
\label{T12}
\end{equation}
with the amplitude for transition between nuclei of $(A,Z)$ and $(A,Z+2)$ accounted for in the matrix element $M^{0\nu}$, and the available phase space for the outgoing electron pair-- along with coupling constants-- expressed as  $G^{0\nu}$. A review of the above, including approaches to the calculation of nuclear matrix elements, is given in Ref.~\refcite{NMEReview2017}. An exploration of possible \onbb~mechanisms, including neutrino exchange, is laid out in Ref.~\refcite{ONBBTheory2012}.

\section{The MAJORANA DEMONSTRATOR}	

The \MJ~\DEM~monitors a mass of \ssGe~in the form of high purity Ge (HPGe) detectors. The detectors are arranged in two compact arrays, with 30\,kg of Ge crystals enriched to 88\% in \ssGe~and an additional 14\,kg of crystals having natural 7.8\% \ssGe~abundance. Background levels are reduced through the use of radiopure materials and insertion of the arrays in a graded passive shield of Cu, Pb, and polyethylene. Additionally, radon gas is purged from the inner shield volume and a surrounding layer of plastic scintillator is used to veto muon events whose flux is reduced through operation of the experiment at the 4850\,ft (4300\,m.w.e.) level of the Sanford Underground Research Facility\cite{SURF2015}.

Background events originating from within the layered shield-- primarily multiply scattered gammas, and alphas at the detector surface-- are rejected through efficient pulse shape cuts. These pulse shape discriminators are enhanced by the novel point-contact detector geometry\cite{LukePPC1989} which localizes the electric potential near the small point-contact electrode; resulting pulse shapes vary strongly depending on the location of ionizing interactions within the detector. 

Two algorithms have been developed to tag and remove the aforementioned gamma and alpha events based on pulse shape. First, a comparison of maximum current $A$ to the energy $E$ of a pulse yields an AvsE parameter that distinguishes multi-site (gamma) events from the single-site events expected for double-beta decay\cite{MJDAvsE2017}. Second, measurements of the decaying edge of a pulse have been developed to flag surface alpha events. Alphas incident on the passivated surface of a detector can ionize charges that slowly drift along that surface and are delayed in their arrival at the electrode. This delayed charge recovery (DCR) counteracts the expected RC decay of a pulse, an effect that is captured by a DCR parameter whose value is used to identify and remove the alpha events\cite{MJDDCR2017,GruszkoThesis2017}. The AvsE and DCR discriminators are tuned to accept 90\% and 99\% of single-site events, respectively, from single-site regions of calibration spectra. High-statistics spectra are produced weekly through insertion of $^{228}$Th calibration sources into the shield\cite{MJDCalibration2017}.

With the above features in mind, the \MJ~Collaboration leverages careful materials selection, a modular design, and the strengths of point-contact HPGe detector technologies to demonstrate the feasibility of a tonne-scale experiment with background levels necessary to explore the inverted ordering region of the \onbb\ parameter space, corresponding to $m_{\beta\beta}$ sensitivities near 20~meV. The Collaboration also takes advantage of low energy thresholds, provided by point-contact detectors, to search for new physics. Initial low-energy analyses have been published\cite{MJDDM2017}, with limits set on bosonic dark matter and solar axion couplings, electron decays, and electronic transitions in violation of the Pauli exclusion principle. Further details of the \MJ~\DEM~experimental design and goals are given in Ref.~\refcite{MJDOverview2014}.

\section{Commissioning and Data Taking}
\label{CommissioningAndDataTaking}

The \MJ~\DEM~began taking commissioning data in June 2015, operating the first cryostat within a partial shield. In December 2015, an inner layer of underground electroformed copper (UGEFCu) was installed around the cryostat cavity, along with other shielding improvements. An additional change replaced the cryostat gaskets-- initially Kalrez-- to lower-activity and lower-mass PTFE. 


Construction of the second cryostat was completed in July 2016 and this additional array of detectors was inserted, alongside the first cryostat, in the cavity at the shield's center. The two cryostats were operated in parallel in August 2016 yielding two data sets whose combination was analyzed for results discussed in Sec.~\ref{RecentResults}. In October 2016, following this parallel run, the independent DAQ systems for each cryostat were merged and data taking resumed. Installation of an outer polyethylene shield was completed in March 2017 and a most recent data set began in May 2017, with both cryostats operating and a full shield in place.

\section{Results}
\label{RecentResults}

\begin{figure}[ht]
	\centerline{\includegraphics[width=8cm]{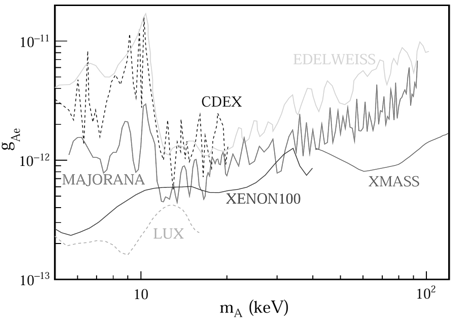}}
	\vspace*{8pt}
	\caption{Experimental limits on the axioelectric coupling of pseudoscalar axionlike dark matter. The curve labeled ``\MJ" is based on 1.3\,kg\,yr of commissioning data with the first cryostat, prior to installation of an inner copper shield. Figure adapted from Ref.~10. \label{gAe_MJDDM2017}} 
\end{figure}

Several physics results have been reported based on subsets of the \DEM's growing data set. Commissioning data were employed for low-energy analyses, setting limits on the axioelectric coupling of pseudoscalar axionlike dark matter, shown in Figure~\ref{gAe_MJDDM2017}, and other beyond-the-Standard-Model processes\cite{MJDDM2017}. With additional data, a \onbb\ half-life sensitivity was calculated based on the first 3.03\,kg\,yr exposure of the first cryostat\cite{MJDElliott2017}.

Soon after this initial exposure, as described in Sec.~\ref{CommissioningAndDataTaking}, both cryostats were operated in parallel. An analysis of this 1.39\,kg\,yr combined exposure, described in Ref~\refcite{MJDGuissepe2017}, estimated the achieved background index to be $1.8^{+3.15}_{-1.14}~\times~10^{-3}$\,c\,(keV\,kg\,yr)$^{-1}$. Since PANIC 2017, analyses of combined data sets have focused on \onbb\cite{MJDDM2017} as discussed in Sec.~\ref{CurrentStatus}.

The Collaboration has also published several technical papers outlining the design of the calibration system\cite{MJDCalibration2017}, results of materials assay\cite{MJDAssay2016}, and results of Ge processing\cite{MJDGeProcessing2018}. A 70\% detector mass yield was attained from the purchased mass of enriched Ge, and cosmogenic activation of $^{68}$Ge and $^{60}$Co in enriched crystals was reduced by a factor of 30 when compared to natural detectors. This reduction in radioactivity within the crystal bulk was achieved through centrifugation, zone refining, and careful shielding during shipment and storage.

The materials assay program employed a number of techniques to constrain the activity of parts used for construction of the experiment. Inductively coupled plasma mass spectrometry (ICPMS) was used to determine activities in \DEM~parts to sub-ppt levels; for instance, U and Th activity in the UGEFCu was limited to $<0.3$\,$\mu$Bq\,kg$^{-1}$. When scaled by the mass of UGEFCu present in the experiment, and by efficiencies determined through detailed Monte Carlo radiation transport simulations, these assay results translate to a 0.23\,c\,(ROI\,t\,yr)$^{-1}$ background contribution\cite{MJDAssay2016}. ROI, here, represents a 4\,keV region of interest surrounding the 2039\,keV $Q$-value of \onbb. Combining simulation-scaled assay results for all other materials and expected background sources, the total expected background rate in the 4\,keV ROI is $<3.5$\,c\,(ROI\,t\,yr)$^{-1}$.

\section{Measuring the \onbb\ Half-life}
\label{CurrentStatus}

Since June 2015, the \DEM~has accrued greater than 20\,kg\,yr of exposure. From this exposure, approximately 10\,kg\,yr were selected for a \onbb\ analysis with a lowest background index of $1.6^{+1.2}_{-1.0}\times10^{-3}$\,c\,(keV\,kg\,yr)$^{-1}$, as detailed in Ref.~\refcite{MJD0nbb2017}. A profile likelihood method was used to determine a \onbb\ half-life and background rate that best fit the data. The likelihood function models the \onbb\ signal as a modified gaussian peak in the presence of a flat background continuum, a shape consistent with the full Monte Carlo simulations.

A lower limit of $1.9\times10^{25}$\,yr was set on the \onbb\ half-life $T^{0\nu}_{1/2}$ through hypothesis testing\cite{MJD0nbb2017}. In this procedure, a series of potential half-life values was proposed, and, for each value, an ensemble of toy Monte Carlo data sets was generated. The aforementioned likelihood methods were then used to determine a best-fit half-life for each toy data set. The lower limit was defined as the largest proposed half-life for which more than 90\% of toy data sets in its ensemble yielded half-lives below the observed best-fit half-life. This limit setting procedure is described in Refs.~\refcite{GERDA0nbb2017}--\refcite{MJDHehn2017}.


\section{Outlook}

With the completion of construction and the achievement of near-background free data in a 360\,keV ROI around the \onbb\ Q-value, the \MJ\ \DEM\ has released its first \onbb\ limit\cite{MJD0nbb2017}. The experiment will continue data taking, with \onbb\ sensitivity projected to increase linearly under the current background conditions. In looking toward next-generation \onbb\ experiments, the development of low-background Ge technology will also continue with planned upgrades to cables and connectors in the \DEM\ cryostats\cite{MJDHaufeReine2017}. 

The achieved milestones of the \DEM\ have established the feasibility of a Ge-based, next-generation \onbb\ search with a much greater detector mass. Feasibility for a Ge-based experiment has also been established by the GERDA Collaboration through an approach complimentary to the \DEM. GERDA has set leading limits on the half-life of \onbb\ in \ssGe\ through operation of Ge crystals in liquid argon, with scintillation light used for active veto\cite{GERDA0nbb2017}. In a next-generation experiment, the most promising technologies of GERDA will be combined with the meticulous materials selection and parts processing developed for the \DEM. The hybrid design has been adopted by the LEGEND Collaboration which incorporates members of both the \MJ\ and GERDA Collaborations along with Ge experts from the international community\cite{LEGENDWilkerson2017}.


\section*{Acknowledgments}

This material is based upon work supported by the U.S. Department of Energy, Office of Science, Office of Nuclear
Physics, the Particle Astrophysics and Nuclear Physics Programs of the National Science Foundation, the Sanford
Underground Research Facility, the Russian Foundation for Basic Research, the Natural Sciences and Engineering Research Council of Canada, the Canada Foundation for Innovation John R. Evans Leaders Fund, the National Energy Research Scientific Computing Center, and the Oak Ridge Leadership Computing Facility.







\end{document}